\documentclass{webofc}

\newcommand{\EEQ}{\end{equation}}
\newcommand{\BEA}{\begin{eqnarray}}
\newcommand{\EEA}{\end{eqnarray}}

\usepackage[varg]{txfonts}   
\woctitle{?????????}
\begin{document}
\title{On the time arrows, and randomness in cosmological signals}
%
%

\author{V.G.Gurzadyan\fnsep\thanks{\email{gurzadyan@yerphi.am}} \and
        S.Sargsyan \and
        G.Yegorian
}
\institute{Center for Cosmology and Astrophysics, Alikhanian National Laboratory and Yerevan State University, Yerevan, Armenia 
          }

\abstract{Arrows of time - thermodynamical, cosmological, electromagnetic, quantum mechanical, psychological - are basic properties of Nature. For a quantum system-bath closed system the de-correlated initial conditions and no-memory (Markovian) dynamics are outlined as necessary conditions for the appearance of the thermodynamical arrow. 
The emergence of the arrow for the system evolving according to non-unitary dynamics due to the presence of the bath, then, is a result of limited observability, and we conjecture the arrow in the observable Universe  as determined by the dark sector acting as a bath. The voids in the large scale matter distribution induce hyperbolicity of the null geodesics, with possible observational consequences.
}
\maketitle
\section{Introduction}
\label{intro}

Time arrows emerge in various processes \cite{P}. The best known is the {\it thermodynamical arrow}, related to the increase of entropy in closed systems.  The expansion of the observed Universe has been associated to the {\it cosmological arrow}. The retarded interaction in electrodynamics defines the {\it electromagnetic arrow}, while the irreversible distortion of the wave function at a 
measurement is defining the {\it quantum mechanical arrow}. The fact that we know the past but not the future, is denoted as a
{\it psychological arrow}. Other processes, such as the black hole collapse or the K$^0$-decay, also reveal time asymmetry. Not all among these arrows are considered to be mutually independent. 

The role of the specific initial conditions for the emergence of the time asymmetry has been outlined by Penrose \cite{P}, while Prigogine \cite {PP} has argued for the importance of the dynamics. The cosmological arrow, according to Page \cite{Page} cannot be explained by inflation but via special (non-correlated) initial conditions. For discussion of various issues related to time arrows and for the sources we refer to e.g. \cite{Hall}, and below we will consider particular issues on the thermodynamical arrow in the context of no-memory dynamics and randomness in cosmological observations.

First, we outline the necessary conditions for the appearance of a thermodynamical arrow based on the system-bath statistical mechanical approach \cite{AG}. Then, the hyperbolicity of the bundle of geodesics induced by the voids, the underdense regions in the large scale matter distribution, along with the observability of the latter process via the study of variations of the randomness in the cosmic microwave background (CMB) maps, will be discussed.

\section{Conditions for the thermodynamical time arrow}   

Considering a quantum system ${\rm S}$ interacting with a thermal bath ${\rm B}$ with an Hamiltonian 
\BEA
\label{mis1}
H=H_{\rm S}+H_{\rm B}+H_{\rm I},
\EEA
where $H_{\rm S}$ and $H_{\rm B}$, $H_{\rm I}$ are their separate Hamiltonians and the interaction part, respectively, one can show (for details see \cite{AG}) that, both the non correlated initial conditions and no-memory, i.e. Markovian dynamics, are necessary conditions to ensure a thermodynamical arrow of a quantum system. 

Important condition is the independence of the states of the bath and the system at initial time 
\BEA
\label{mis3}
{\cal D}(0)=D_{\rm S}(0)\otimes D_{\rm B}(0),
\EEA
where the density matrix ${\cal D}(t)$ of the system satisfies von Neumann equation
\BEA
\label{mis2}
i\partial _t{\cal D}(t) =[H,{\cal D}(t)],\quad
{\cal D}(t)=e^{-itH/\hbar}{\cal D}(0)e^{itH/\hbar}.
\EEA

Then the state of the system at time $t$ is given by the partial density matrix
\BEA
\label{rashid}
D_{\rm S}(t)={\rm tr}_{\rm B}{\cal D}(t),
\EEA
where ${\rm tr}_{\rm B}$ denotes the trace over the Hilbert space of the 
bath. 

The system-bath approach implies the fact of limited observability. The system and
the bath form a closed system, where one deals with the state of the
system only. The latter under the influence of the bath evolves via a 
non-unitary dynamics determined by a superoperator ${\cal T}$
\BEA
\label{khalil}
D_{\rm S}(t)={\cal T}(t,0)\,D(0)=
\sum_{\alpha\beta}A_{\alpha\beta}D(0)A^\dagger_{\alpha\beta},
\EEA
where $A_{\alpha\beta}$ are operators in Hilbert space of ${\rm S}$. 

Within this approach we conjecture  {\it the emergence of the arrow in the observable Universe due to the dark sector acting as a bath}. According to current estimations, the dark sector includes the dark energy (73\%) and dark matter (23\%). While the dark matter is assumed to have usual thermodynamical potentials, the dark energy is accelerating the expansion of the Universe possessing negative pressure. 

The superoperator ${\cal T}$ in (\ref{khalil}) is not
unitary, and does not have an inverse operator, i.e. the dynamics of the quantum system alone is irreversible. 
As mentioned above, this time asymmetry is a consequence of the fact that
statistical systems are described incompletely.  The dynamics of the system is autonomous,
determined by the initial condition (\ref{mis3}), although this
property is not valid for arbitrary initial state.

The property 
\BEA
{\cal T}(t_{f},t_i)={\cal T}(t_{f}-t_i)
\EEA
at $t_f> t_i$, which is automatically valid for 
the unitary situation, is violated as the presence of the bath \cite{AG}.

Eqs.~(\ref{mis2}, \ref{mis3}) are enough to ensure the existence of the pre-arrow, i.e. the one not surviving during the entire dynamics of the system but only indicating the similarity between the processes given by (\ref{khalil}) and described by the same 
Hamiltonian (\ref{mis1}) but at somewhat different initial condition  
\BEA
\label{mis4}
{\cal R}(0)=R_{\rm S}(0)\otimes D_{\rm B}(0).
\EEA
To ensure the transformation of the pre-arrow to an arrow, dynamical conditions
are required. At sufficiently large $t$ the initial conditions have to be of no role:
\BEA
\label{baba}
{\cal T}(t,0)={\cal T}(t),
\EEA
and the system has to tend to a certain stationary density matrix 
$D^{(st)}_{\rm S}$:
\BEA
\label{handi}
{\cal T}(t)D^{(st)}_{\rm S}=D^{(st)}_{\rm S},\quad
D_{\rm S}(t)\to D^{(st)}_{\rm S}.
\EEA
The form of the thermodynamical potential depends on $D^{(st)}_{\rm S}$, e.g. if the stationary distribution is microcanonical
$D^{(st)}_{\rm S}\propto 1$, then one will have the increase with time of the von Neumann entropy 
\BEA
\label{s1}
S_{vN}[D_{\rm S}(t)]=-{\rm tr}[D_{\rm S}(t)\ln
D_{\rm S}(t)].
\EEA 

Thus, in system-bath approach we discussed the possibility when the de-correlated initial conditions and no-memory dynamics emerge a thermodynamical arrow in the observed Universe with the dark energy acting as a bath.  

Note that for a weak coupling with the bath, the relaxation consists of
two different steps: during the first step the average energy exchange
with the bath is not essential, and the relaxation is microcanonical:
the system gets equal probability to be in all states with the same
(initial) energy. Clearly, the entropy increases here with the same
mechanism as in (\ref{s1}). During the second step, the energy exchange
becomes essential. In the nuclear magnetic resonance/electron spin 
resonance (NMR/ESR) physics these two steps are known as,
respectively, $T_2$ and $T_1$ processes \cite{oja}. 

Finally, we should stress that the dark energy has an unusual
thermodynamic feature: negative pressure. It is presently an open
question to which extent this feature will (or will not) lead to
thermodynamic pathologies, or whether the thermodynamic description can
be reformulated by making the negative pressure analogous to a negative
temperature (both are thermodynamic potentials); see e.g. \cite{braz}.
Note that negative temperature systems were realized experimentally \cite{oja},
and it is known that the negative temperature can be accomodated to the
structure of thermodynamics \cite{oja}. However, we should stress that
within the weak system-bath coupling regime, these aspects will influence only the
second relaxation regime, where the exchange of energy is essential. The
first step is largely independent on the issues related to negative
temperature and negative pressure. 

\section{Voids and hyperbolicity}

Now we will consider a mechanism of no-memory dynamics due to the matter distribution inhomogeneities in the large-scale Universe \cite{GK}.
For perturbed Friedmann-Robertson-Walker metric
\BEA
ds^2 =  -(1+2\phi)\ dt^2 + (1-2\phi)\ a^2(t)\ \gamma_{mn}(x)dx^mdx^n\ ,\label{PRWMetric}
\EEA
with perturbation $|\phi|\ll 1$, the averaged Jacobi equation, i.e. the geodesic deviation one, for the length of the deviation vector $\ell$ 
can be written in the form 
\BEA
\frac{d^2\ell}{d\lambda^2}+ r\ \ell = 0\ ,
\EEA
where
\BEA
\lambda(z,\Omega_\Lambda,\Omega_m)
=\int_0^z\frac{d\xi}{\sqrt{\Omega_\Lambda +[1- \Omega_\Lambda + \Omega_m \xi]\ (1+\xi)^2}}\
\EEA
and
\BEA
r &=& -\Omega_k +2\Omega_m\delta_0\ ,\\
\delta_0&\equiv&\frac{\delta\rho_0}{\rho_0}\ ,
\EEA
$\Omega_i, i=k, m, \Lambda$ are density parameters for the curvature, matter (luminous and dark) and dark energy, respectively.
 
For $\Omega_k=0$ and $\Omega_\Lambda +\Omega_m = 1$, one has
\BEA\label{elltau}
\frac{d^2\ell}{d\tau^2}+ \delta_0\ \ell = 0\ ,
\EEA
where
\BEA
\tau(z,\Omega_m)&=&\sqrt{2\Omega_m}\ \lambda(z,1-\Omega_m,\Omega_m)\\
&=&\sqrt{2\Omega_m}\int_0^z\frac{d\xi}{\sqrt{1+\Omega_m[(1+\xi)^3-1],}}\ 
\EEA
i.e. even in the globally flat $k=0$ FRW Universe, the voids with negative density contrast $\delta_0$ will induce
hyperbolicity of null geodesics. The large scale galaxy surveys give for the mean density contrast
of the voids the values $-(0.8-0.9)$ \cite{Ho,Su}. 

If the density contrast $\delta_0$ is periodic $\delta_0(\tau+\tau_\kappa+\tau_\omega)=\delta_0(\tau)$ \cite{GK} and
\BEA
\delta_0=
\begin{cases}
+\kappa^2\ & 0<\tau<\tau_k\ ,\\
-\omega^2\ & \tau_\kappa<\tau<\tau_\kappa+\tau_\omega\ ,
\end{cases}
\EEA
then the solution of the Jacobi equation is unstable if (e.g. \cite{Arnold})
\BEA
\mu=\left|2\cos(\omega\tau_\omega)\cosh(\kappa\tau_\kappa)
+\left(\frac{\kappa}{\omega}-\frac{\omega}{\kappa}\right)
\sin(\omega\tau_\omega)\sinh(\kappa\tau_\kappa)\right|-2>0\ .
\EEA
In our case $0<\omega\tau_\omega\ll 1$ and $0<\kappa\tau_\kappa\ll 1$. It is easy to see that if 
\BEA
\bar{\delta_0}&\equiv&\int_0^{\tau_\kappa+\tau_\omega}\delta_0 d\tau
=\kappa^2\tau_\kappa-\omega^2\tau_\omega\ne 0\ ,
\EEA
then
\BEA
\mu\sim \bar{\delta_0}(\tau_\kappa + \tau_\omega)\ ,
\EEA
and $\mu<0$, if $\bar{\delta_0}<0$, and $\mu>0$, if $\bar{\delta_0}>0$. On the other hand, if $\bar{\delta_0}=0$ then one can show that
\BEA
\mu\sim -\frac{[(\kappa\tau_\kappa)^2+(\omega\tau_\omega)^2]^2}{12}<0\ .
\EEA

The solutions are stable, if $\bar\delta_0\le 0$ and unstable, if $\bar\delta_0>0$. In terms of physical values for the mean densities of the voids and the walls and their sizes $L$ one has
\BEA
\bar\delta_0\sim -\delta_{void}L_{void}-\delta_{wall} L_{wall} \ .
\EEA
For the voids $\delta_{void} \simeq -1$, and the instability condition above yields
\BEA
L_{void}>\delta_{wall} L_{wall} \ .
\EEA
which can be written in the form
\BEA
p>\frac{\delta_{wall}}{\delta_{wall}+1}, 
\EEA
where 
\BEA
p=L_{void}/(L_{wall}+L_{void})\le 1\ 
\EEA
is denoting the porosity in the matter distribution.  

The corresponding Lyapunov exponent $\chi$ is 
\BEA
\chi = \log s\ ,
\EEA
where
\BEA
s = 1+\frac{\mu}{2}+\sqrt{\left(1+\frac{\mu}{2}\right)^2-1}\sim e^{\sqrt{\mu}}\ ,
\EEA
for $\mu\ll 1$. Therefore, $\chi\sim\sqrt{\mu}$ , and the mixing rate $b\sim s^n\sim e^{n\chi}$, where $n=\tau/(\tau_\kappa +\tau_\omega)$. Thus,
\BEA
n\chi\sim n\sqrt{\bar{\delta_0}(\tau_\kappa + \tau_\omega)}\sim n\sqrt{p-\delta_{wall}(1-p)}\ \left(\frac{L_{wall}+L_{void}}{L}\right)
=\sqrt{p-\delta_{wall}(1-p)}\ ,
\EEA
and finally
\BEA
b\sim e^{\sqrt{p-\delta_{wall}(1-p)}}\ .
\EEA

\section{Degree of randomness}

Here we will consider the degree of randomness as defined by Kolmogorov \cite{Kolm} and Arnold \cite{Arnold_N,Arnold_UMN,Arnold_MMS} applied to a cosmological signal, i.e. to cosmic microwave background datasets. 

For a sequence of real-valued variables the stochasticity parameter is defined 
via the theoretical $F(x)$ and empirical $F_n(x)$ distribution functions  
\begin{equation}\label{KSP}
\lambda_n=\sqrt{n}\ \sup_x|F_n(x)-F(x)|\ .
\end{equation}

Kolmogorov theorem \cite{Kolm} states that for any continuous $F$ the probability limit is
\begin{equation}
\lim_{n\to\infty}P\{\lambda_n\le\lambda\}=\Phi(\lambda)\ ,
\end{equation}
where 
\begin{equation}
\Phi(\lambda)=\sum_{k=-\infty}^{+\infty}\ (-1)^k\ e^{-2k^2\lambda^2}\ ,\ \  \lambda>0\ , \Phi(0)=0 \label{Phi}
\end{equation}
the convergence is uniform and Kolmogorov's distribution $\Phi$ is independent on $F$.
 
The stochasticity parameter $\lambda_n$ is a random quantity and the interval of its probable values is determined by the distribution function $\Phi$: $0.3\le\lambda_n\le 2.4$. 

Consider a class of sequences
\begin{equation}
z_n = \alpha x_n + (1-\alpha) y_n,
\end{equation}
where $x_n$ and $y_n$ are random and regular sequences, respectively, $\alpha$ varying within $[0,1]$ and indicating the fraction of those sequences. For
\begin{equation}
y_n = \frac{an\pmod b}{b},
\end{equation}
where $a$ and $b$ are mutually fixed prime numbers, numerical experiments have revealed the behavior of the stochasticity parameter \cite{GGS}, see e.g. (Figure 1).

\begin{figure}
\includegraphics[width=6cm,clip]{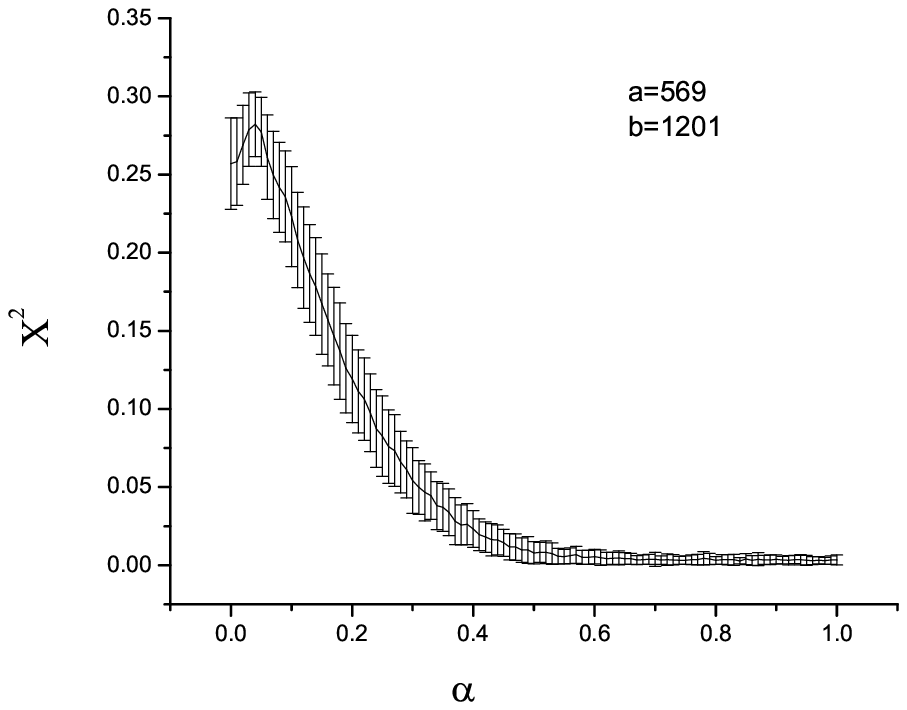}
\includegraphics[width=7cm,clip]{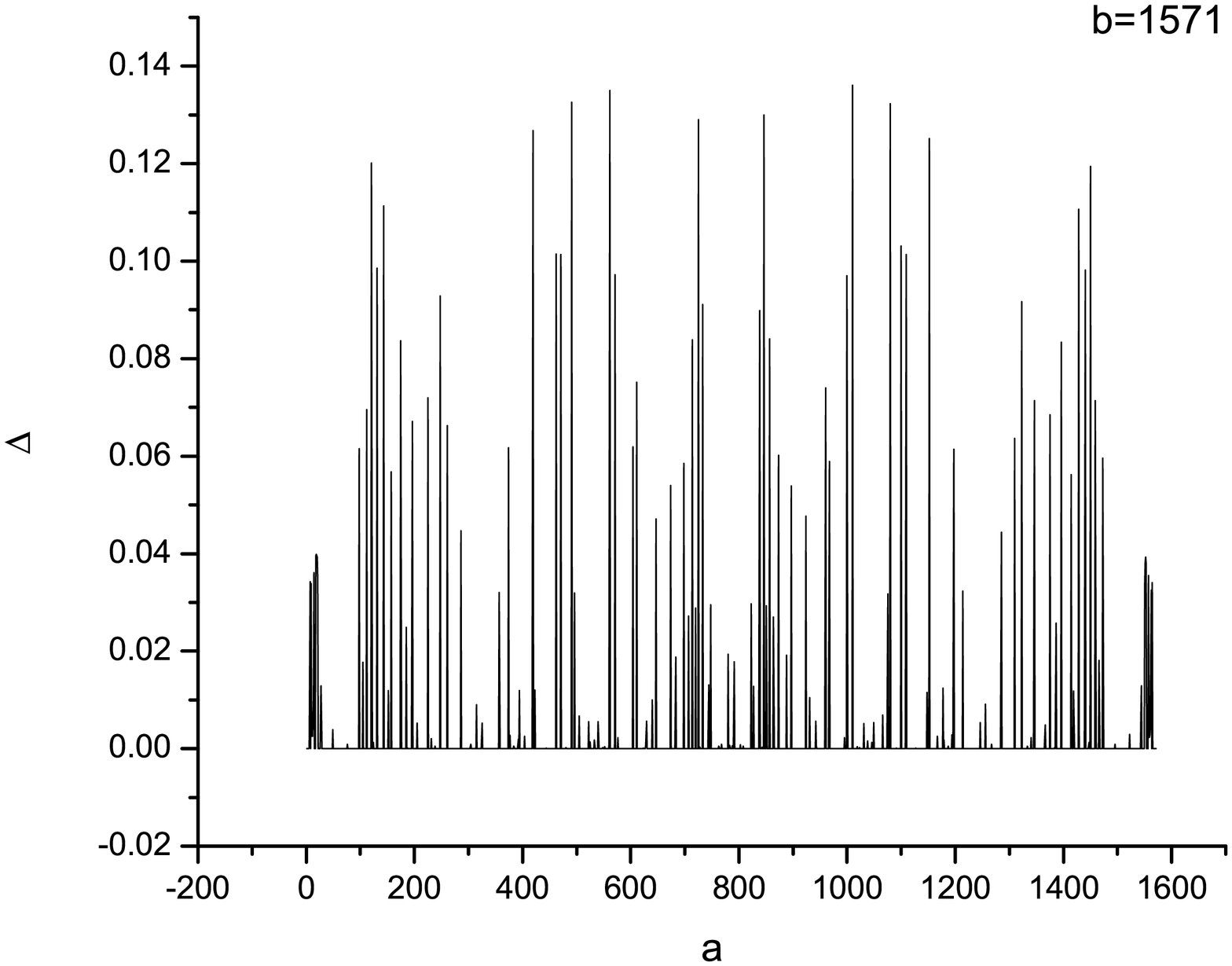}
\caption{The $\chi^2$ for the function $\Phi$ and a uniform one for the sequence $z_n$ vs $\alpha$ indicating the contribution of random and regular subsequences for a pair of input parameters (left plot). A scaling in the $\chi^2$ vs the input parameter of the regular subsequence (right plot; see \cite{GGS}).}
\label{fig-1a}       
\end{figure}

\begin{figure}
\includegraphics[width=5cm,clip]{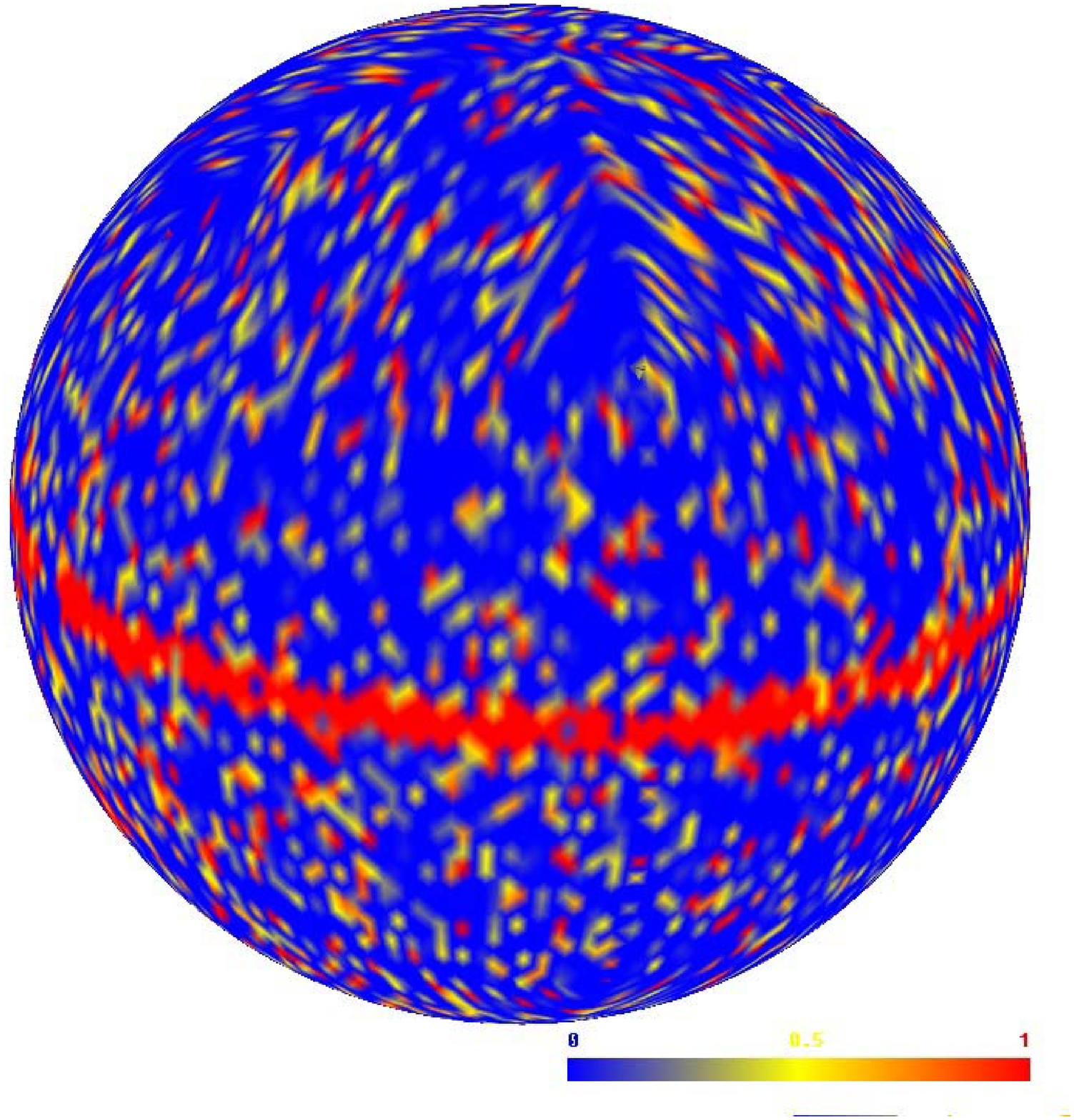}
\includegraphics[width=6cm,clip]{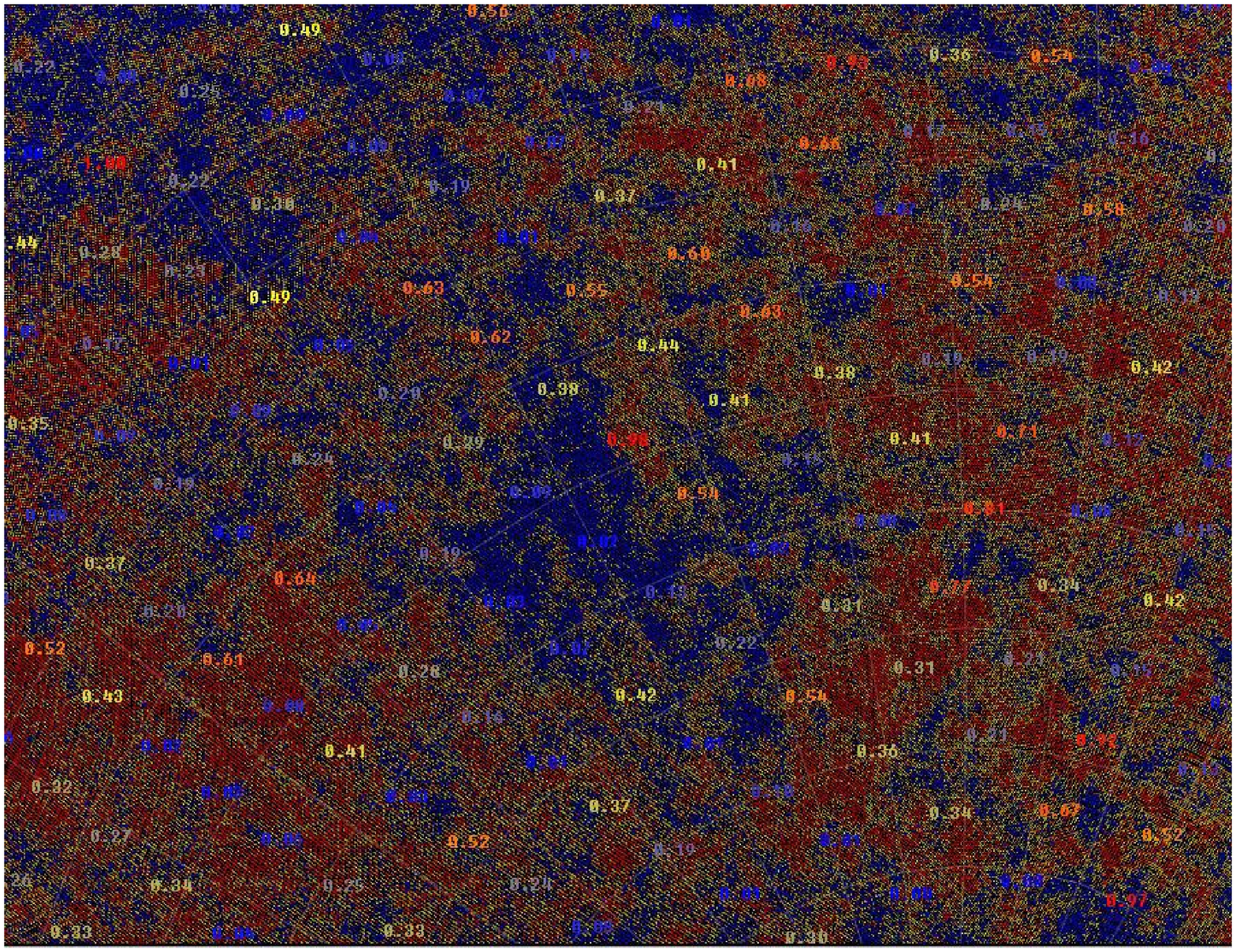}
\caption{The sky map of the function $\Phi$ obtained for Wilkinson Microwave Anisotropy 
Probe's 7-year cosmic microwave background temperature dataset: full sky (left plot) ; a sky region sample with indicated values of  $\Phi$ (right plot).}
\label{fig-3}       
\end{figure}

The function $\Phi(\lambda)$ i.e. the degree of randomness has been obtained for the CMB datasets of Wilkinson Microwave Anisotropy 
Probe \cite{G1,G2} (Figure 2).  The resulting sky map enables to separate the cosmological signal (CMB) from non-cosmological ones, e.g. the Galactic disk, and also point sources (quasars, blazars), including those not detected by other methods. 

The non-Gaussian region of the CMB sky, known as a Cold Spot \cite{VM}, reveals enhanced degree of randomness \cite{G2}. If it is due to the hyperbolicity of the voids discussed above, then it would support the void nature of the Cold Spot, as argued by some authors (e.g. \cite{In}). Other regions of higher randomness, although not as clearly outlined as the Cold Spot, were also revealed in the sky.  

The degree of randomness of CMB, therefore, can be helpful not only for the separation of cosmological and non-cosmological signals but also for probing the large scale matter distribution.

The conjecture on the emergence of the thermodynamical arrow of the observed Universe due to the dark sector acting as a bath, can become a subject of further analysis.

We thank A.E.Allahverdyan and A.A.Kocharyan for valuable discussions and help.

\end{document}